\documentclass[aps,twocolumn,showpacs,amsmath,amssymb]{revtex4}

\usepackage{graphicx}

\newcommand{\eone}{\hat{E}_1}
\newcommand{\eoned}{\hat{E}_1^{\dagger}}

\newcommand{\wone}{\hat{\Omega}_1}

\newcommand{\woned}{\hat{\Omega}_1^{\dagger}}

\newcommand{\etwo}{\hat{E}_2}
\newcommand{\etwod}{\hat{E}_2^{\dagger}}

\newcommand{\wtwo}{\hat{\Omega}_2}
\newcommand{\wtwod}{\hat{\Omega}_2^{\dagger}}

\newcommand{\be}{\begin{eqnarray}}
\newcommand{\ee}{\end{eqnarray}}

\begin{document} 

\title{Full quantum solutions to the resonant four-wave mixing 
of two single-photon wave packets} 
\author{Mattias Johnsson and Michael Fleischhauer} 
\affiliation{Fachbereich Physik, Universit\"{a}t Kaiserslautern, D-67663 Kaiserslautern, 
Germany} 
\date{\today} 
 
\begin{abstract} 
We analyze both analytically and numerically the resonant four-wave mixing
of two co-propagating single-photon wave packets. We present analytic
expressions for the two-photon wave function and
show that soliton-type quantum solutions exist which display a 
shape-preserving oscillatory exchange of excitations between the modes. 
Potential applications including quantum information processing are
discussed.
\end{abstract} 
 
 
\maketitle 
 



\section{Introduction}


The cancellation of resonant linear absorption and refraction 
via electromagnetically induced transparency (EIT) 
\cite{harris1997,Marangos} has led to a range of new possibilities in
non-linear optics. 
One important application is
optical frequency mixing close to atomic resonances 
which allows the use of the enhanced nonlinear 
interaction without suffering from linear absorption and refraction. 
It has been predicted that EIT could lead to a new regime of nonlinear 
optics on the level of few light quanta 
\cite{Harris1998,HauHarris99,Imamoglu96,Lukin-AdvAMO-2000,johnsson2002a}. 
Several schemes for resonant nonlinear processes have been
proposed and analyzed, both theoretically and experimentally
\cite{Lukin-AdvAMO-2000}. 
A particularly interesting system is resonant four-wave mixing using 
atoms with a double-lambda configuration
\cite{Hemmer1994,Babin1996,Popov1997,Lu1998}.
Efficient frequency conversion, generation of squeezing 
\cite{LukinPRL1999}
as well as 
the possibility of mirrorless oscillations 
\cite{Lukin1998}
with extremely low thresholds 
and narrow linewidth have been predicted 
\cite{Fleischhauer-PRL-1999} and in part experimentally
observed \cite{Zibrov-PRL-2000}.

Most theoretical and experimental studies of resonant nonlinear processes
have been done for classical fields. For one-dimensional set-ups 
where common co-moving frames exist, full analytical solutions have
been derived for the interaction of classical pulses in the adiabatic
limit \cite{Korsunsky2002}. In view of the potential 
for an efficient nonlinear interaction on the level of few photons, however,
a full quantum theoretical analysis of these systems is necessary.
In addition, in order to take into account finite size effects which become
increasingly important in the few-photon regime, and to 
analyze the potential for quantum information processing, such a quantum
analysis has to go beyond the usual linearization approaches.
Here very little work has been done, the few exceptions being the 
integrable models of resonantly enhanced Kerr interaction 
\cite{Imamoglu-Kerr} and photon blockade \cite{Harris1998,blockade}.

In a previous paper \cite{johnsson2002a}
we have shown numerically that if single-mode fields are
considered, it is possible to use an atomic vapor in a 
four-wave mixing double-lambda configuration to obtain
full conversion from two input fields into two generated fields within
a few centimeters of interaction length, even if the input fields only 
consist of single light quanta. The treatment was fully quantum
but being restricted to a single-mode analysis, propagation
effects of wave packets were not considered.
In the present paper we extend this study by
undertaking a multimode analysis in one spatial dimension and for
co-propagating fields. In order to keep the problem tractable we restrict 
ourselves to the important special case of two  single-photon wave packets
as inputs. For this case we obtain simple analytic solutions 
and compare them to numerical simulations.
We show that soliton-like quantum solutions exist that display an
oscillatory exchange of excitation between the two input and the
two generated fields. Finally we discuss
briefly possible applications of single-photon four-wave mixing to
quantum information processing and entanglement generation.


\section{System and effective field equations}

\begin{figure}[ht] 
  \begin{center} 
    \includegraphics[width=5cm]{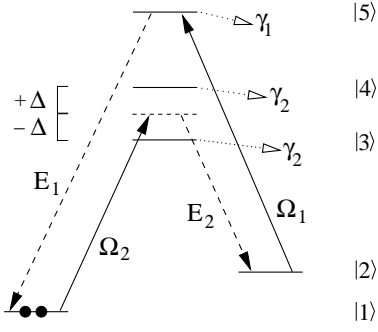} 
    \caption{Four-wave mixing in a modified double-$\Lambda$ system
with sgn[$d_{42}/d_{41}]=-$sgn$[d_{32}/d_{31}]$, with $d_{ij}$ being the
dipole moment of the $|i\rangle - |j\rangle$ transition.}
    \label{fig5level} 
  \end{center} 
\end{figure}

The situation we consider is resonantly enhanced four-wave mixing in the
modified double-lambda system shown in Fig.~\ref{fig5level}. Note
that a five-level atomic system is used instead of the original four-level
system put forward in \cite{Hemmer1994,Lukin1998}. 
This is due to the fact that in the four-level system the finite
detuning $\Delta$ is associated with an AC Stark effect, which
leads to intensity dependent dynamical phase shifts of the
fields. These phase shifts are of minor consequence in the case where
the fields are counter-propagating \cite{Fleischhauer00b}, but for
co-propagating fields have a detrimental influence leading to impaired
phase matching and inefficient frequency conversion. 
As shown in \cite{johnsson2002b} these phase shifts
are eliminated in the five-level scheme when the relative sign of the
dipole moments of the 
$|4\rangle \to |2\rangle, |1\rangle$ transitions is opposite
to that of  the $|3\rangle \to |2\rangle, |1\rangle$ transitions.


The two fields with carrier frequencies $\nu_{\Omega 1}$ and $\nu_{\Omega 2}$
and slowly varying amplitudes 
$\Omega_1$ and $\Omega_2$ are initially excited and form
the pump fields. The other fields of carrier frequencies 
$\nu_{E1}$ and $\nu_{E2}$ and slowly varying amplitudes
$E_1$ and $E_2$  are generated during the interaction
process and are assumed to be initially zero. 
$\Omega_1$ and $E_1$ are taken to be exactly on resonance,
i.e. $\nu_{\Omega 1}=\omega_{52}$, $\nu_{E_1}=\omega_{51}$,
while the other two fields are detuned by an amount
$\Delta$, i.e. $\nu_{\Omega 2}=\omega_{41}-\Delta=\omega_{31}+\Delta$
and $\nu_{E2}=\omega_{42}-\Delta=\omega_{32}+\Delta$. 
A finite detuning $\Delta$, large compared to the Rabi
frequencies, the Doppler broadening and the decay rates from the excited
states, is necessary to maximize the ratio of nonlinear gain to linear
absorption. Decay from the two lower levels is considered to be
negligible and
all fields have the same propagation direction.

Because of energy conservation there is overall
four-photon resonance, i.e. $\nu_{\Omega 1}+\nu_{\Omega 2}=
\nu_{E1}+\nu_{E2}$. It can be shown 
that the contributions of the resonant transitions 
to the linear refractive index vanish if the fields 
are pairwise in two-photon resonance. Phase matching will thus favor 
two-photon resonance and we assume that this condition is fulfilled
for the carrier frequencies of the four pulses, i.e.
$\nu_{E1}-\nu_{\Omega 1}=\nu_{\Omega 2}-\nu_{E2}=\omega_{21}$.

Extending the analysis of \cite{johnsson2002a} to a multimode description, 
the interaction can be described by the effective adiabatic Hamiltonian
\cite{hamiltonian}
\be
H_{\rm int} = \frac{\hbar g c}{\Delta}
\int\!\! {\rm d}z\, \frac{\hat\Omega_1^\dagger\hat\Omega_2^\dagger \hat E_1\hat E_2
+\hat E_1^\dagger\hat E_2^\dagger \hat\Omega_1\hat\Omega_2}
{\hat\Omega_1^\dagger\hat\Omega_1+\hat E_1^\dagger\hat E_1}.
\label{Hint}
\ee
Here $\hat\Omega_1(z),\hat\Omega_1^\dagger(z)$ 
etc. denote dimensionless, slowly-varying (both in time and space) 
positive and negative-frequency components of the corresponding
electric fields
\be 
\hat E_j(z) &=& \frac{1}{\sqrt{L}}\sum_{k} \hat a_{jk}\, {\rm e}^{ikz}
\, {\rm e}^{i\nu_{Ej}(t-z/c)} \label{eqEdef}
,\\
\hat \Omega_j(z) &=& \frac{1}{\sqrt{L}}\sum_{k} \hat b_{jk}\, {\rm e}^{ikz}
\, {\rm e}^{i\nu_{\Omega j}(t-z/c)}. \label{eqWdef}
\ee
$L$ is the quantization length and $k\ge -\nu/c$ is the wave-vector component
in $z$-direction relative to $\nu/c$. 
The four fields are assumed to have either sufficiently
different carrier frequencies or different polarizations. Thus operators
corresponding to different fields commute. The commutator between 
positive and negative frequency components can be approximated
by a spatial delta function
\be
[\hat E_i(z),\hat E_j^\dagger(z^\prime)] &=& 
\frac{\delta _{ij}}{L}\, \sum_{k} {\rm e}^{ik(z-z^\prime)} \approx 
\delta_{ij}\, \delta(z-z^\prime).\label{commuteE}
\ee
We furthermore assume in (\ref{Hint}) that all four transitions give rise to
the same coupling $g=N d_i^2 \omega_i /(2\hbar c\epsilon_0)= 
3 N \lambda^2 \gamma/8 \pi$, where $N$ is the atomic number density,
$\lambda$ the typical wavelength of the fields and $\gamma$ the typical 
radiative decay rate. 
It should be noted that numerator
and denominator of Eq. (\ref{Hint}) commute and thus there is
no ambiguity with respect to the ordering of the terms. 

The structure of the denominator
results from the saturation of the
two-photon transition $|1\rangle - |2\rangle$, whose coherence lifetime
is taken to be infinite. If a finite decay rate $\gamma_0$ 
of the $|1\rangle - |2\rangle$ coherence is taken into account, a term
proportional to $\gamma\gamma_0$ has to be added in the denominator.

The non-polynomial character of the interaction Hamiltonian 
causes the nonlinear coupling to behave unusually: As shown
in \cite{johnsson2002a}, the interaction increases with decreasing pump
field strength, making effective nonlinear frequency conversion possible
even for single photons. In the derivation of
the effective Hamiltonian in \cite{johnsson2002a} adiabatic conditions were 
assumed. This limits the applicability of (\ref{Hint}) in the
multimode case to sufficiently long pulses. A discussion of non-adiabatic
corrections and their effect on the propagation of the pulses is,
however, outside of the scope of the present paper and will be discussed
elsewhere. 

As shown in the Appendix the slowly varying amplitudes of the electric field
obey
\be
\partial_t\, \hat E_j(z,t) = -c \, \partial_z \, \hat E_j(z,t)
+\frac{i}{\hbar}\bigl[\hat H_{\rm int},\hat E_j(z,t)\bigr]
\label{eqGeneralEOM}
\ee
and similarly for $\hat\Omega_j$.
Thus from (\ref{Hint}) we arrive at the following equations of motion
\begin{eqnarray}
\Bigl(\partial_t + c\,\partial_z\Bigr) \eone 
&=& i\kappa c \hat{\Lambda} \Bigl(\woned \wtwod \eone \eone
\etwo - \woned \etwod \wone \wone \wtwo\Bigr) \hat{\Lambda} \nonumber \label{eqEOM1} \\
\Bigl(\partial_t + c\,\partial_z\Bigr) \wone &=& i\kappa c 
\hat{\Lambda} \Bigl(\eoned \etwod \wone \wone
\wtwo - \eoned \wtwod \eone \eone \etwo\Bigr) \hat{\Lambda} \nonumber 
\label{eqEOM2} \\
\Bigl(\partial_t + c\, \partial_z\Bigr) \etwo &=& -i\kappa c \hat{\Lambda}
\eoned \wone \wtwo \nonumber \label{eqEOM3} \\ 
\Bigl(\partial_t + c\, \partial_z\Bigr) \wtwo &=& -i\kappa c \hat{\Lambda} \woned \eone \etwo \label{eqEOM4}
\end{eqnarray}
where $\kappa=g / \Delta$ and $\hat{\Lambda} = (\woned\wone +
\eoned\eone)^{-1}$.

These equations admit four independent constants of motion: 
\begin{eqnarray}
(\partial_t + c\, \partial_z )(\hat \Omega_1^\dagger \hat \Omega_1 + \hat E_1^\dagger\hat E_1) &=&
0 \nonumber \label{eqCOM1} \\
(\partial_t + c\, \partial_z) (\hat \Omega_2^\dagger \hat \Omega_2 + \hat E_2^\dagger\hat E_2) &=&
 0 \nonumber \label{eqCOM2} \\
(\partial_t + c\, \partial_z) (\hat \Omega_1^\dagger \hat \Omega_1
-\hat \Omega_2^\dagger \hat \Omega_2) &=& 0 \nonumber \label{eqCOM3} \\
(\partial_t + c\, \partial_z) (\hat \Omega_1^\dagger \hat
\Omega_2^{\dagger} \hat E_1 \hat E_2  + \hat \Omega_1 \hat \Omega_2
\hat E_1^{\dagger} \hat E_2^{\dagger}) &=& 0, \label{eqCOM4}
\end{eqnarray}
which represent the quantum analogs of the Manley-Rowe relations and
additionally the conservation of the relative phase between the fields.

If we assume that the input fields consist of two, single-photon
wave packets in  $\wone$ 
and $\wtwo$, then it is clear, due to the constants of motion (\ref{eqCOM4}), that
the state of the system can be represented at all times by
\begin{equation}
|\varphi(t)\rangle = \sum_{k,k'}\xi_{kk'}(t) |1_k \, 1_{k'} \, 0 \, 0
\rangle + \sum_{k,k'}\eta_{kk'}(t) |0 \, 0 \, 1_k \, 1_{k'}
\rangle, \label{eqgeneralstatevector}
\end{equation}
where $|n_k\, m_{k^\prime} \, p_{k^{\prime\prime}}\, q_{k^{\prime\prime\prime}}
\rangle$ denotes $n$ photons in the $k$th mode of $\hat\Omega_1$,
$m$ photons in the $k^\prime$th mode of $\hat\Omega_2$, and so on.


\section{Field intensities}


Since for the case of a single-photon input the expectation values
of all fields vanish at all times, all relevant information about the state
of the system is given by the mean intensities of the fields
\be
\langle\varphi(t)\vert \hat\Omega_j^\dagger(z)\hat\Omega_j(z)
|\varphi(t)\rangle,
\qquad
\langle\varphi(t)\vert \hat E_j^\dagger(z)\hat E_j(z)|\varphi(t)\rangle,
\ee
and the two-photon wave functions
\begin{eqnarray}
\psi_\Omega(z,z',t) &=& \langle 0 \,|\, \wone (z) \wtwo (z') \,|\,
\varphi(t) \rangle \nonumber \\
%
%
&=& \sum_{k,k'=-\infty}^{\infty} e^{2\pi ikz/L}\, e^{2\pi ik'z'/L}
\,\xi_{kk'}(t) \label{eq2DFT}\\
\psi_E(z,z',t) &=& \langle 0 \,|\, \eone (z) \etwo (z') \,|\,
\varphi(t)\rangle \nonumber \\
%
%
&=& \sum_{k,k'=-\infty}^{\infty} e^{2\pi ikz/L}\, e^{2\pi ik'z'/L}
\,\eta_{kk'}(t). \label{eq2DFT-e}
\end{eqnarray}
$\psi_\Omega(z,z^\prime,t)$ and
$\psi_E(z,z^\prime,t)$ represent the amplitude
of finding the $\wone$ ($\eone$) photon at
position $z$ and simultaneously the $\wtwo$ ($\etwo$) 
photon at position $z'$. 
The $\psi$'s are 2D Fourier transforms from the
$k$-space representations $\xi_{kk^\prime}$ and $\eta_{kk^\prime}$
into a real-space representation. 

We first discuss the dynamics of the mean intensities of the fields.
Due to the constants of motion it is sufficient to calculate say $\langle
\hat\Omega_1^\dagger \hat\Omega_1\rangle$.
\begin{eqnarray}
\Bigl(\partial_t + && \!\!\!\!\!\! c \partial_z\Bigr) \langle \varphi
|\hat{\Omega}_1^{\dagger}
  \hat{\Omega}_1 | \varphi\rangle =
 \langle \varphi |\frac{i}{\hbar} \left[
\hat{H},\hat{\Omega}_1^{\dagger}\hat{\Omega}_1
  \right] | \varphi \rangle \nonumber \\
&&= i\kappa c \, \langle \Bigl(\hat{\Omega}_1^{\dagger} \hat{\Omega}_1
\hat{\Omega}_1^{\dagger}
\hat{\Omega}_2^{\dagger} \hat{E}_1 \hat{E}_2 - \hat{\Omega}_1 \hat{\Omega}_2
\hat{E}_1^{\dagger} \hat{E}_2^{\dagger} \hat{\Omega}_1^{\dagger}
 \hat{\Omega}_2^{\dagger}\Bigr) \hat{\Lambda} \rangle \nonumber \\
&&= i\kappa c\, \langle \hat{E}_1^{\dagger} \hat{E}_2^{\dagger} \hat{\Omega}_1
\hat{\Omega}_2 - \hat{\Omega}_1^{\dagger} \hat{\Omega}_2^{\dagger} \hat{E}_1
\hat{E}_2 \rangle
\end{eqnarray}
where we have dropped the common spatial coordinate $z$ and used
$\hat{H}|0\rangle=0$, $\hat{\Lambda}|\varphi\rangle = |\varphi\rangle$ as
well as the fact that three annihilation operators acting on
$|\varphi\rangle$ give zero. Similarly we find
\begin{eqnarray}
\Bigl( \partial_t +c \partial_z \Bigr) && \!\!\!\!\!\! \langle
\hat{E}_1^{\dagger}
 \hat{E}_2^{\dagger} \hat{\Omega}_1 \hat{\Omega}_2 - \hat{\Omega}_1^{\dagger}
  \hat{\Omega}_2^{\dagger} \hat{E}_1 \hat{E}_2 \rangle \nonumber \\
&&= 2i\kappa c \langle \hat{\Omega}_1^{\dagger} \hat{\Omega}_2^{\dagger}
 \hat{\Omega}_1 \hat{\Omega}_2 - \hat{E}_1^{\dagger} \hat{E}_2^{\dagger}
    \hat{E}_1 \hat{E}_2 \rangle \\
\Bigl( \partial_t +c \partial_z \Bigr) &&  \!\!\!\!\!\! \langle
\hat{\Omega}_1^{\dagger}
 \hat{\Omega}_2^{\dagger}  \hat{\Omega}_1 \hat{\Omega}_2 -
\hat{E}_1^{\dagger}
   \hat{E}_2^{\dagger} \hat{E}_1 \hat{E}_2 \rangle \nonumber \\
&&= 2i\kappa c \langle \hat{E}_1^{\dagger} \hat{E}_2^{\dagger} \hat{\Omega}_1
  \hat{\Omega}_2 - \hat{\Omega}_1^{\dagger} \hat{\Omega}_2^{\dagger} \hat{E}_1
    \hat{E}_2 \rangle.
\end{eqnarray}
Consequently the differential equation we must solve is
\begin{equation}
\Bigl( \partial_t +c \partial_z \Bigr)^3 \langle \hat{\Omega}_1^{\dagger}
  \hat{\Omega}_1\rangle = -4\kappa^2 c^2 \Bigl( \partial_t +c \partial_z
 \Bigr) \langle \hat{\Omega}_1^{\dagger} \hat{\Omega}_1 \rangle.
\end{equation}
We take the input to be two independent single-photon wave packets
in $\hat\Omega_1$ and $\hat\Omega_2$ with the same spatial envelope
$f_0(z)$ and vacuum in the other two fields, which corresponds 
to a separable initial state of the form
\be
|\varphi\rangle_{\rm in} &=& \Bigl[\sum_{k}\, \xi_k^0 |1_k\rangle_{\Omega_1}\Bigr]
\otimes \Bigl[\sum_{k}\, \xi_k^0 |1_k\rangle_{\Omega_2}
\Bigr] \nonumber \\
&& \otimes \bigl|0\bigr\rangle_{E1}\otimes \bigl|0\bigr\rangle_{E2}.
\label{init-state}
\ee
The $\xi_k^0$ are Fourier-transforms of $f_0(z)$
\be
f_0(z)=\sum_k \xi_k^0 \, {\rm e}^{2\pi i k z/L}.
\ee
Thus
\be
\langle \hat\Omega_1^\dagger(z,t)\hat\Omega_1(z,t)\rangle_{\rm in}
&=&\langle \hat\Omega_2^\dagger(z,t)\hat\Omega_2(z,t)\rangle_{\rm in}
\nonumber\\
&=&\psi_0(z-ct)
\ee
where $\psi_0(z)=f^2_0(z)$. With these initial conditions one finds
\begin{equation}
\langle \hat{\Omega}_1^{\dagger} \hat{\Omega}_1 \rangle =
\langle \hat{\Omega}_2^{\dagger} \hat{\Omega}_2 \rangle =
\psi_0(z-ct) \cos^2 (\kappa z),
\end{equation}
i.e. a sinusoidal exchange of excitation between the two
pump and the two generated fields. Complete conversion is
achieved at $z=\pi/\kappa$. It is worthwhile noting that 
as shown in \cite{johnsson2002a}, and contrary to the classical
dynamics, complete conversion can only be achieved
in the quantum case for initial Fock states with one or two
photons in the two pump modes.


\section{Dynamics of the two-photon wave function; quantum solitons}

\begin{figure}[ht] 
  \begin{center} 
    \includegraphics[width=8cm]{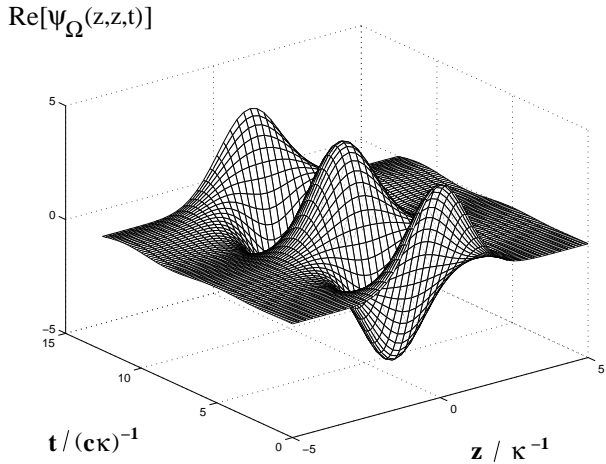} 
    \caption{45-mode simulation of wave function propagation in a
co-moving frame. Both initial wave packets taken to be Gaussian.} 
    \label{fig1} 
  \end{center} 
\end{figure}

We proceed by calculating the two-photon wave functions. 
The calculation of $\psi_\Omega(z,z',t)$ can be split into two distinct
cases, depending on whether $z=z'$ or $z\ne z^\prime$. 
Let us first assume $z=z'$. 
\begin{eqnarray}
\Bigl(\partial_t + && \!\!\!\!\!\! c\, \partial_z\Bigr) 
\langle 0 \,| \wone \wtwo \, 
| \, \varphi \rangle =
\langle 0 \, | \, \frac{i}{\hbar} \left[ \hat{H},\wone \wtwo\right] | \,
\varphi\rangle \nonumber \\ 
&& = -i \kappa c \langle 0 \,| \, \wone \wtwo \Bigl(\woned
\wtwod \eone \etwo + \wone \wtwo \eoned \etwod\Bigr) 
\hat{\Lambda} \,| \, \varphi \rangle \nonumber \\
&& = -i \kappa c \langle 0 | \eone\etwo | \psi \rangle
\end{eqnarray}
where we have again used the results that  
$\hat{H}|0\rangle =0$, $\hat{\Lambda}|\varphi\rangle=|\varphi \rangle$ and
that three annihilation operators acting on $|\varphi\rangle$ give zero. 
Operating on $\langle 0 | \eone\etwo | \psi \rangle$ we find 
\begin{equation}
(\partial_t + c \, \partial_z) \langle 0 \,| \eone \etwo \, 
| \, \varphi \rangle =
-i \kappa \langle 0 \, | \, \wone \wtwo \, | \, \varphi\rangle. \\
\end{equation}
Thus, to determine the wave function, we need to solve the differential equation
\begin{equation}
\Bigl(\partial_t + c\, \partial_z \Bigr)^2 \psi_\Omega(z,z,t) = -\kappa^2
c^2 \psi_\Omega(z,z,t).
\label{eqDEforz=z}
\end{equation}
For an input consisting of two independent single-photon wave packets
with the same spatial shape, as considered above, the initial two-photon
wave function reads
\be
\psi_\Omega(z,z^\prime,t)_{\rm in} = f(z-ct)f(z^\prime-ct).
\ee
We also have 
at the entrance of the medium
\begin{equation}
\Bigl(\partial_t + c\, \partial_z\Bigr) \psi_\Omega(z,z,t)\Bigr|_{z=0} 
=  -i \kappa c \langle 0
| \eone\etwo | \varphi \rangle\Bigr|_{z=0} = 0
\end{equation}
so that the solution to (\ref{eqDEforz=z}) is given by
\begin{equation}
\psi_\Omega(z,z,t) = \psi_0(z-ct) \cos(\kappa z),
\label{eqWwf}
\end{equation}
with $\psi_0(z) = f^2(z)$.
Thus the two-photon wave function at equal spatial points
propagates through the medium
modulated by a factor of $\cos(\kappa z)$. 
We see that after one full conversion cycle
$z=\pi/\kappa$ the phase of the wave function has changed sign. This
agrees with a numerical simulation of the quantum problem, the results
of which are shown in Fig.~\ref{fig1}.
%
%


%
%

\begin{figure}[ht] 
  \begin{center} 
    \includegraphics[width=8cm]{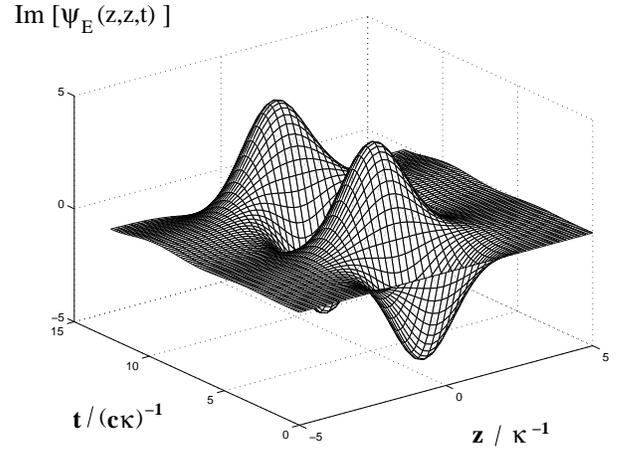} 
    \caption{Generated fields in 45-mode simulation. Both initial
wave packets taken to be Gaussian.} 
    \label{fig2} 
  \end{center} 
\end{figure}

We can use a similar procedure to find $\psi_E(z,z,t) = \langle 0 |\eone\etwo
|\varphi\rangle$, the wave function of the generated fields. We obtain
\begin{equation}
\psi_{E}(z,z,t) = -i \,\psi_0 (z-ct) \sin(\kappa z).
\label{eqEwf}
\end{equation}
The evolution of this field is shown in Fig.~\ref{fig2}. Note that it
is $\pi/2$ out of phase with the drive field wave functions, as expected.
%
%
 
%
%

\

\begin{figure}[ht] 
  \begin{center} 
    \includegraphics[width=8cm]{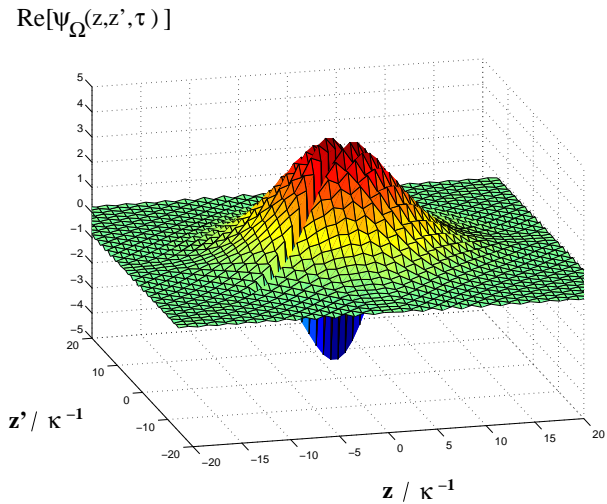} 
    \caption{Numerical 45-mode calculation of $\psi_\Omega(z,z',\tau)$, shown
after one full cycle of energy transferal from pump to generated
fields and back again. The sign flip for $z=z^\prime$ is evident.
} 
    \label{fig2Dwf} 
  \end{center} 
\end{figure}

For full knowledge of the state of the system, we also need to find
$\psi(z,z',t)$ where $z\neq z'$. From (\ref{commuteE})
we see that operators at different points in space commute, so that
\begin{eqnarray}
\Bigl(\partial_t + && \!\!\!\!\!\!\! \, c\, \partial_z+ c\, 
\partial_{z^\prime}
\Bigr)\langle 0 |
\wone(z)\wtwo(z') | \varphi\rangle \\
&& \hspace{-5mm}=\frac{i}{\hbar} \langle 0| \left[\hat{H},\wone (z)\right] 
\wtwo(z')
+ \wone(z) \left[\hat H,\hat\Omega_2(z')\right] |\varphi\rangle\nonumber
\end{eqnarray}
Expanding the Hamiltonian, normal ordering the expression and again
noting the form of (\ref{eqgeneralstatevector}) we eventually arrive at
\begin{equation}
\Bigl(\partial_t + c\, \partial_z + c\, \partial_{z'}\Bigr) \, \langle 0 |
\wone(z)\wtwo(z') | \varphi\rangle = 0. 
\label{eqDEfor2Dwf}
\end{equation}
Thus the wave function of the system in the case where $z \neq z'$ is
simply given by the corresponding input expression
\begin{equation}
\psi(z,z',t) = f(z-ct)f(z^\prime-ct), \hspace{0.5 cm} z\neq z'.
\end{equation}
It is evident that the initial two-photon wave function at different 
coordinates propagates undisturbed throughout
the medium, and that there is a discontinuous change in the
behavior
when moving away from the line $z=z'$. 
This is clearly seen in Fig.~\ref{fig2Dwf},
which shows the two dimensional wave function $\psi_\Omega(z,z',\tau)$ at time
$\tau=\pi/\kappa$. This is the time required for one full conversion
from the pump fields to the generated fields and back again.
Essentially we have at this time
\begin{eqnarray}
\psi_\Omega(z,z',\tau) = -\psi_\Omega(z,z')_{\rm in}
 && \hspace{5mm} z=z' \label{eqsignflip} \\
\psi_\Omega(z,z',\tau) = \psi_\Omega(z,z')_{\rm in} && \hspace{5mm} z\neq z'
\label{eqnosignflip} 
\end{eqnarray}
%

This behavior can easily be understood. The two-photon wave function
represents the joint probability density that two photons are located
at positions $z$ and $z^\prime$. As both photons propagate at the same
speed they will never meet at the same point in space unless they did
initially. 
Since the nonlinear interaction in (\ref{Hint})
is local, two photons at different spatial points do not
interact. Thus only if $z=z^\prime$ is there a conversion from the
pump fields into the generated fields due to
the local nonlinearity. 

The simple sign change in the wave function for $z=z'$ hints at the
possibility of using the system as a phase gate for quantum computation,
as was mentioned in the single-mode case considered in
\cite{johnsson2002a}. One chooses the length of the nonlinear medium
such that exactly one full conversion cycle can occur. If the entire
wave function changes sign, we would have a system that behaves as a true phase
gate: if only one of the two inputs is populated, the pulse exits the
medium unchanged, but if both are present, a phase shift of $\pi$ occurs.

However, due to the multimode nature of this
problem, it is $\psi(z,z',t)$ for all $z$ and $z^\prime$ 
that is a true reflection of the system
rather than $\psi(z,z,t)$ and there is no sign change
for $z\ne z^\prime$. To see whether we can still use the
system as an approximate phase gate, we consider the behavior of the 
state vector coefficients in
(\ref{eqgeneralstatevector}). Returning briefly to a picture with $2N+1$
discrete modes, by using Eq. (\ref{eqlocaloperatordef}) of the Appendix
one can show that
(\ref{eqsignflip}) and (\ref{eqnosignflip}) imply  
\begin{equation}
\xi_{kk'}(\tau) = \xi_{kk'}(0) - \frac{2}{2N+1}\sum_{mn} \delta_{m+n,k+k'}\,
\xi_{mn}(0). \label{eqxicondition}
\end{equation} 
This gives a good picture of how the different modes have mixed among
themselves, and shows that in general a simple sign change of all the
coefficients cannot exist. Only if we can enforce that at least temporarily 
only a single effective mode of the two pump fields is excited, e.g. by using
a resonator set-up is it possible to use the system as a phase gate.

We see however from (\ref{eqxicondition})
that after one full conversion cycle 
the initially factorized state (\ref{init-state}) evolves into
an entangled state between all modes of the fields $\hat\Omega_1$
and $\hat \Omega_2$. Thus the nonlinear interaction generates 
entanglement.  

On the other hand if the initial state is a two-photon wave packet
in an entangled state, such that the initial two-photon wavefunction has
only a contribution for $z=z^\prime$
\be
\psi_\Omega(z,z^\prime)\Bigr \vert_{\rm in}
=\phi_0(z)\, \delta(z-z^\prime)
\ee
only diagonal components of the two-photon wave function
will ever be nonzero. According to (\ref{eqWwf}) and (\ref{eqEwf}) they 
undergo sinusoidal oscillations
\be
\psi_\Omega(z,t) &=& \phi_0(z-ct)\, \cos(\kappa z),\\
\psi_E(z,t) &=& -i\phi_0(z-ct)\, \sin(\kappa z).
\ee
The superposition of pump and generated fields
\be
\Phi(z,t)&\equiv& \cos(\kappa z)\psi_\Omega(z,t)+i\sin(\kappa z)\, \psi_E(z,t)
\nonumber\\
&=&\phi_0(z-ct)
\ee
propagates in a form-stable manner and represents a quantum soliton solution.
The two-photon wavefunction $\Phi(z,t)$ of the quantum soliton corresponds
to a quasi-particle excitation 
\be
&&\hat\Psi^\dagger(z,t) \equiv\\
&&\quad\cos(\kappa z)\hat\Omega_1^\dagger(z,t)\hat\Omega_2^\dagger(z,t)
-i\sin(\kappa z) \hat E_1^\dagger(z,t)\hat E_2^\dagger(z,t).\nonumber
\ee


\section{summary}


In the present paper we have presented a full quantum theoretical
treatment of resonant forward four-wave mixing for pulses. For this
we have used an effective Hamiltonian derived from the interaction
of the fields with an ensemble of atoms in a double-lambda configuration
using an adiabatic approximation \cite{johnsson2002a}. We were particularly
interested in the few-photon regime since here quantum effects dominate
the dynamics and because of potential applications in quantum information
processing. Thus we have restricted
ourselves to the important special case of an input
consisting of two single-photon wave packets. For this case we were able
to analytically solve the propagation equations for the field intensities
and two-photon wave functions which contain all relevant information
about the quantum state. 

We found that there is an oscillatory energy exchange 
between the two pump and generated fields with 100\% conversion at 
periodic intervals of interaction. This result is characteristic for a 
few-photon Fock-state input; for a coherent input complete conversion
can only be achieved asymptotically for very large input power. 

We have also shown that after even multiples of the conversion length 
the two-photon wave function $\psi(z,z^\prime,t)$ regains its initial 
form, while after odd multiples there is a sign flip 
for $z=z^\prime$. 

If the two input wave packets are not independent but in a highly
entangled state, the two-photon wave function can be made zero outside
of the diagonal. It was shown that such a pair of input wave packets 
form a formstable soliton-like quantum solution which is a superposition
of pump and generated fields with oscillating coefficients. 

The process of resonant four-wave mixing was shown to generate
large entanglement between the modes forming the two, single-photon
wave packets. Furthermore the nonlinear interaction strength is
large enough to generate a controlled phase shift of a single photon
by the presence of another one. Thus, if the number of relevant
modes is at least temporarily restricted by some external means like
a resonator, the system could have interesting
applications as a photonic phase gate. 


\section*{Acknowledgement}


The authors acknowledge financial support by the Deutsche
Forschungsgemeinschaft through
the special program on quantum information and the Graduiertenkolleg
792 at the University of Kaiserslautern.


\section*{Appendix}


To simplify the transition from a single-mode description to a
multimode one, we will 
first consider one single-mode quantum field $\hat{a}$, and then
generalize the result to our four-field system.
Suppose that the
single-mode Hamiltonian governing the evolution of $\hat{a}$ is given by
$\hat{H}=\hbar \omega_0 \, \hat{a}^{\dagger} \hat{a} +
\hat{H}_{\rm int}(\hat{a},\hat{a}^{\dagger})$. To go over to the
multimode description we
consider an interaction region of length $L$, divided into
$2N+1$ cells, and consider a discrete set of modes around the carrier
frequency of the field, i.e. $k_n = k_0 + 2n\pi /L$, $-N\leq n \leq N$. 
We now define localized field operators (denoted by a tilde) via
\begin{eqnarray}
\hat{\tilde{a}}_{l} &=& \sum_{k=-N}^{N}\hat{a}_{k}
\exp\left[\frac{2\pi ikl}{2N+1}\right]
\label{eqlocaloperatordef} \\
\hat{a}_{k} &=& \frac{1}{2N+1}\sum_{l=-N}^{N}\hat{\tilde{a}}_{l}
\exp\left[\frac{-2\pi ikl}{2N+1}\right]
\end{eqnarray}
where the $\hat{a}_k$ are annihilation operators for mode
$k$. $\hat{\tilde{a}}_{l}$ is related to the field strength in
cell $l$.

These localized field operators have the commutation relations
\begin{equation}
\Bigl[ \hat{\tilde{a}}_{l}, \hat{\tilde{a}}_{l'}^{\dagger}\Bigr] = (2N+1)\delta_{l l'}.
\label{eqcommutationrelations}
\end{equation}

\noindent The multimode Hamiltonian can now be written as
\begin{eqnarray}
\hat{H} &=& \hbar \sum_{k}\omega_k \hat{a}_{k}^{\dagger} \hat{a}_{k} +
\sum_{l} \hat{H}_{\rm int}(\hat{\tilde{a}}_l,\hat{\tilde{a}}_l^{\dagger})  \\ 
&=& \frac{\hbar\omega_0}{2N+1} \sum_l \hat{\tilde{a}}_{l}^{\dagger}
\hat{\tilde{a}}_{l} - \hbar\sum_{l l'}\omega_{l l'} \hat{\tilde{a}}_{l}^{\dagger}
\hat{\tilde{a}}_{l'} + \sum_{l}
\hat{H}_{\rm int}(\hat{\tilde{a}}_l,\hat{\tilde{a}}_l^{\dagger})
\nonumber  
\end{eqnarray} 
where $\omega_0$ is the carrier frequency and
\begin{equation}
\omega_{l l'} = \sum_{k=-N}^N \frac{2\pi kc}{(2N+1)^2 L} \exp\left[
\frac{2\pi i k (l-l')}{2N+1}\right].
\end{equation}

Commutation relations yield the following Heisenberg equation of motion
\begin{equation}
\frac{\partial}{\partial t} \hat{\tilde{a}}_{l} = -i \omega_0
\hat{\tilde{a}}_{l} \, -i(2N+1)\sum_{l'} \omega_{l l'} \hat{\tilde{a}}_{l'}
+\frac{i}{\hbar} [\hat{\tilde{H}}_{\rm int},\hat{\tilde{a}} ] 
\end{equation}
Now, as we let $N\rightarrow\infty$ we find 
\begin{eqnarray}
lL/(2N+1) & \rightarrow & z \\
\hat{\tilde{a}}_{l}  & \rightarrow & \hat{a}(z,t) \\
-i(2N+1)\sum_{l'} \omega_{l l'} \hat{\tilde{a}}_{l'}
& \rightarrow & -c \, \partial_z \hat{a}(z,t) \\
\left[ \hat{a}(z,t), \hat{a}^{\dagger}(z',t)\right] & \rightarrow & L \, \delta(z-z').
\end{eqnarray}
Introducing slowly time-varying amplitudes we obtain
\begin{equation}
\left(\partial_t + c\, \partial_z \right) \hat{a}(z,t) = \frac{i}{\hbar}
[\hat{H}_{\rm int}(t),\hat{a}(z,t) ] 
\label{eq_example_multimodeEOM}
\end{equation}
Thus the multimode equations of motion
look exactly like the single-mode equations of motion with the
exception that a $c\,\partial_z$ term has been added and the fields now
have a spatial dependence.

Returning to the four-wave mixing situation described in this paper, we see
that the interaction Hamiltonian given by (\ref{Hint}) is of the
form shown above, with the summation replaced by an integral in the continuum
limit. The multimode annihilation and creation operators 
$\hat{E}(z)$ and $\hat{\Omega}(z)$ defined in (\ref{eqEdef}) and 
(\ref{eqWdef}) are analogous to the localized field operators $\hat{a}(z,t)$
defined above, except for the factor of $\sqrt{L}$ inserted to ensure correct
commutation relations regardless of the quantization length. Thus the equation 
of motion (\ref{eqGeneralEOM}) follows from 
(\ref{eq_example_multimodeEOM}) above.

\def\etal{\textit{et al.}}

\end{document}